\documentclass[5p,10pt,a4paper]{elsarticle}

\usepackage{amsmath}
\usepackage{amssymb}
\usepackage[USenglish]{babel}
\usepackage{color}
\usepackage{float}
\usepackage{graphicx}
\usepackage{hyperref}
\usepackage{listings}
\usepackage{mathtools}
\usepackage{microtype}
\usepackage{physics}

\allowdisplaybreaks

\newcommand\ee{\mathrm{e}}
\newcommand\ii{\mathrm{i}}
\newcommand\bigO{\mathcal{O}}

\newcommand\mbcrea[1]{\mathrm{#1}^\dagger}
\newcommand\mbannh[1]{\mathrm{#1}^{\mathchoice{\vphantom\dagger}{}{}{}}}
\newcommand\cee{\mbannh{c}}
\newcommand\cdag{\mbcrea{c}}
\newcommand\eff{\mbannh{f}}
\newcommand\fdag{\mbcrea{f}}

\newcommand\iw{\ii\omega}

\newcommand\wmax{\bgroup\omega_{\mathrm{max}}\egroup}
\newcommand\TT{\mathcal T}
\newcommand\WW{\mathcal W}

\newcommand\sparseir{{\em sparse-ir}}

\definecolor{purple}{rgb}{0.6, 0.2, 0.8}
\definecolor{orange}{rgb}{0.91, 0.41, 0.17}

\newcommand\printarg[1]{\if\relax\detokenize{#1}\relax\else: #1\fi}

\restylefloat{table}
\journal{SoftwareX}

\begin{document}

\begin{frontmatter}

% ==========================
\title{sparse-ir: optimal compression and sparse sampling of many-body propagators}
% ==========================

\author[tuwien]{Markus Wallerberger}

% alphabetic for now - order TBD
\author[tuwien]{Samuel Badr}
\author[saitama]{Shintaro Hoshino}
\author[saitama]{Fumiya Kakizawa}
\author[tohoku]{Takashi Koretsune}
\author[CCSE,aip]{Yuki Nagai}
\author[kyoto]{Kosuke Nogaki}
\author[rcast]{Takuya Nomoto}
\author[cems]{Hitoshi Mori}
\author[okayama]{Junya Otsuki}
\author[deptphys]{Soshun Ozaki}
\author[saitama]{Rihito Sakurai}
\author[tuwien]{Constanze Vogel}
\author[hamburg]{Niklas Witt}
\author[issp]{Kazuyoshi Yoshimi}

\author[saitama,presto]{Hiroshi Shinaoka}

\address[tuwien]{Department of Solid State Physics, TU Wien, 1040 Vienna, Austria}
\address[saitama]{Department of Physics, Saitama University, Saitama 338-8570, Japan}
\address[tohoku]{Department of Physics, Toh\=oku University, Miyagi 980-8577, Japan}
\address[kyoto]{Department of Physics, Kyoto University, Kyoto 606-8502, Japan}
\address[okayama]{Institute for Interdisciplinary Science, Okayama University, Okayama 700-8530, Japan}
\address[deptphys]{Department of Physics, University of Tokyo, Bunkyo, Tokyo 113-0033, Japan}
\address[rcast]{Research Center for Advanced Science and Technology, University of Tokyo, 4-6-1 Meguro-ku, Tokyo, 153-8904, Japan}
\address[cems]{RIKEN Center for Emergent Matter Science (RIKEN CEMS) 2-1 Hirosawa, Wako, Saitama 351-0198, Japan}
\address[hamburg]{I. Institute of Theoretical Physics, University of Hamburg, 22607 Hamburg, Germany}
\address[issp]{Institute for Solid State Physics, University of Tokyo, Tokyo 113-8654, Japan}
\address[CCSE]{CCSE, Japan Atomic Energy Agency, Kashiwa, Chiba 277-0871, Japan}
\address[aip]{Mathematical Science Team, RIKEN Center for Advanced Intelligence Project (AIP), Tokyo 103-0027, Japan}
\address[presto]{JST, PRESTO, 4-1-8 Honcho, Kawaguchi, Saitama 332-0012, Japan}
  
\begin{abstract}
We introduce sparse-ir, a collection of libraries to efficiently handle imaginary-time propagators, a central object in finite-temperature quantum many-body calculations.
We leverage two concepts: firstly, the intermediate representation (IR), an optimal compression of the propagator with robust {\em a-priori} error estimates, and secondly, sparse sampling, near-optimal grids in imaginary time and imaginary frequency from which the propagator can be reconstructed and on which diagrammatic equations can be solved.
IR and sparse sampling are packaged into stand-alone, easy-to-use Python, Julia and Fortran libraries, which can readily be included into existing software.  We also include an extensive set of sample codes showcasing the library for typical many-body and {\em ab initio} methods.
\end{abstract}


\end{frontmatter}

\section*{Code metadata}

{\footnotesize
\noindent\begin{tabular}{|l|p{\dimexpr.5\columnwidth-2.6em\relax}|p{\dimexpr.5\columnwidth-2.6em\relax}|}
\hline
\textbf{Nr.} & \textbf{Code metadata description} & \textbf{Please fill in this column} \\ \hline
C1 & Current code version
& 1.0-beta1\\
\hline
C2 & Permanent link to code/repository used for this code version
&
\href{https://github.com/SpM-lab/sparse-ir}{github.com/SpM-lab/sparse-ir};
\href{https://github.com/SpM-lab/SparseIR.jl}{github.com/SpM-lab/SparseIR.jl};
\href{https://github.com/SpM-lab/sparse-ir-fortran}{github.com/SpM-lab/sparse-ir-fortran}
\\ 
\hline
C3 & Code Ocean compute capsule & 
%\TODO{}
--
\\
\hline
C4 & Legal Code License   & MIT \\
\hline
C5 & Code versioning system used & git \\
\hline
C6 & Software code languages, tools, and services used & Python or Julia or Fortran \\
\hline
C7 & Dependencies & 
scipy (optional: xprec)
\\
\hline
C8 & Link to developer documentation/manual &
\href{https://sparse-ir.readthedocs.io}{sparse-ir.readthedocs.io};
\href{https://spm-lab.github.io/sparse-ir-tutorial}{spm-lab.github.io/sparse-ir-tutorial}
\\
\hline
C9 & Support email for questions & 
\href{https://github.com/SpM-lab/sparse-ir/issues}{github.com/SpM-lab/sparse-ir/issues}
\\
\hline
\end{tabular}
}
\vfill

%\linenumbers
% ============================
\section{Motivation and significance}
\label{sec:motivation}
% ============================
Computational quantum many-body physics is a major driver of advances in materials
science, quantum computing, and high-energy physics.  Yet, in pushing these fields forward,
we face a three-pronged challenge: firstly, the requirement to model
more complicated systems in an effort to understand advanced many-body effects,
secondly, speeding up the calculations to allow large-scale automatized system
discovery, and thirdly, the need for reliable error control to fortify predictive power
of the results.

For diagrammatic methods working in imaginary (Euclidean) time---widely used to solve quantum
many-body systems---these three prongs translate to the need to compactly store, quickly manipulate,
and reliably control the error, respectively, of many-body propagators and the diagrammatic equations in which they
appear.  Previous efforts either focused on optimizing imaginary time grids~\cite{Wei02,Kananenka16} or
modelling generic smooth functions~\cite{Boehnke11,Dong20}.

The intermediate representation (IR)~\cite{Otsuki17,Shinaoka17:compressing} instead leverages
the analytical structure of imaginary-time propagators to construct a maximally compact, orthonormal basis:
the number of basis functions needed to represent a propagator scales logarithmically with the desired 
accuracy and logarithmically with $\Lambda$, the ultraviolet cutoff in units of temperature.
(Related approaches either optimize for a
different norm~\cite{Kaltak20} or trade some compactness for simpler algorithms \cite{DLR21,libdlr}.)
Sparse sampling~\cite{Li20} is a complementary concept which connects the IR to
sparse time and frequency grids, which allows us to efficiently move between representations
and restrict
the solution of diagrammatic equations to those grids.
Uniquely, error control is baked into the IR: each basis function comes with
an {\em a priori} error level, which also means changing accuracy is simply a matter of changing the number of nonzero basis
coefficients.   The IR for the one-particle basis also serves as a building block for compressing arbitrary
$n$-point propagators~\cite{Shinaoka18:overcomplete,Shinaoka20:tensornw} and fast solutions to
the corresponding diagrammatic equations~\cite{Wallerberger21:BSE}.

Precomputed IRs for different cutoffs $\Lambda$ have been released previously as the
{\em irbasis} library~\cite{irbasis}.  Using this library, IR and sparse sampling has 
been successfully employed in numerous physics and chemistry applications 
~\cite{NomotoPRB2020,NomotoPRL2020,NomuraRR2020,Isakov2020,Niklas2020,Pokhilko2021-uv,Yeh2021-fu,Ye2022,witt2022doping,Nagai:2019dea,nagai_quasicrystal,Itou21,sakurai2021,Nagai2022}.

In the paper, we introduce \sparseir{}, a major step forward from the previous library: it computes the
basis on the fly, usually within seconds. This not only removes the need for precomputing and shipping
databases, it also allows tayloring the cutoff $\Lambda$ and even the type of kernel to the specific
application. We also simplify the use of sparse sampling, which previously had to be implemented
on top of {\em irbasis} by the user.  Finally, we improve the infrastructure for two-particle calculations
by adding the possibility of augmented and vertex bases~\cite{Shinaoka18:overcomplete,Wallerberger21:BSE}.
We also provide a set of small, self-contained Jupyter notebooks showcasing the use of IR and sparse
sampling for selected physics and quantum chemistry applications, lower the barrier of entry for new users.
The library is available as three standalone Python, Julia and Fortran ports, each with minimal dependencies.

The remainder of this paper is organized as follows: after an overview over IR and sparse sampling in Sec.~\ref{sec:software}, we showcase the use of \sparseir{} in a simple Feynman diagrammatic method in
Sec.~\ref{sec:examples}.  In Sec.~\ref{sec:arch}, we then give an overview of the anatonomy and function
of the package. We state our final assessments in Sec.~\ref{sec:impact}.

% =============================
\section{Software description}
\label{sec:software}
% =============================
We are concerned with (retarded) many-body propagators and related functions in equilibrium:
\begin{equation}
    G^\mathrm{R}(\omega) = -\ii\int_{t'}^\infty \dd{t} \ee^{\iw t} \langle A(t) B(t') \mp B(t') A(t) \rangle,
\end{equation}
where both $A, B$ are bosonic ($-$) or fermionic ($+$) operators, $\langle\cdot \rangle = \Tr(\ee^{-\beta H} \; \cdot\,)/\Tr(\ee^{-\beta H})$ is the expectation value, $1/\beta$ is temperature and $H$ is the Hamiltonian.
At its core, \sparseir{} seeks to (i) maximally compress the information contained in these propagators and (ii) reliably reconstruct this
compressed form from sparse time and frequency grids to allow its use in diagrammatic calculations.

\begin{figure}
    \centering
    \includegraphics[width=\columnwidth]{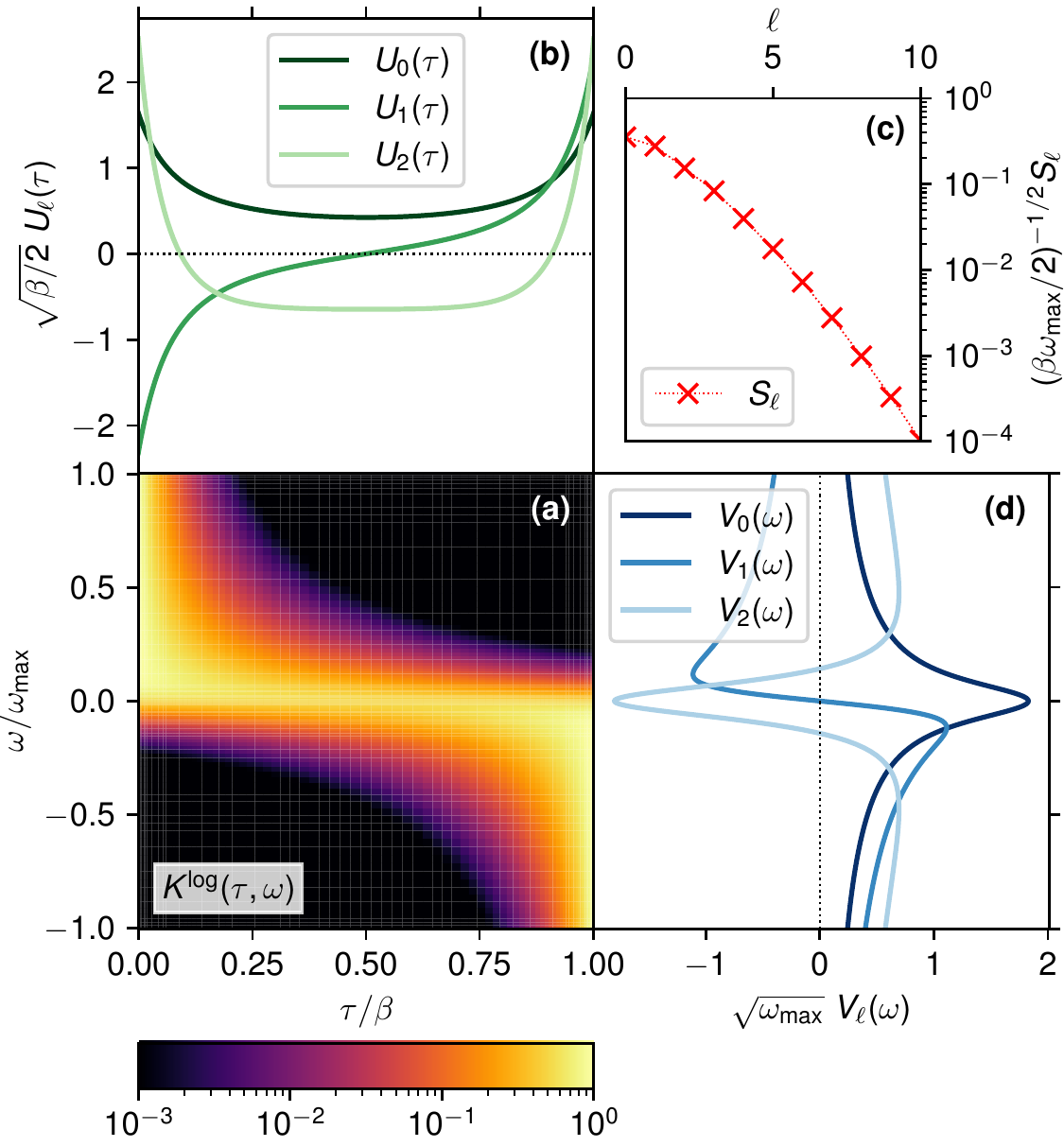}
    \caption{Singular value expansion (\ref{eq:SVE}) of the (a) analytic continuation kernel
        $K$ (\ref{eq:K}) for $\beta\wmax=30$ into (b) left-singular functions $U_l$ on the 
        imaginary-time axis, (c) singular values $S_l$, and (d) right-singular functions $V_l$
        on the real-frequency axis.}
    \label{fig:sve}
\end{figure}

To achieve (i) compression, \sparseir{} relies on the fact that information is lost in
transitioning from the (observable) spectral function $\rho(\omega)=-\frac1\pi\Im G^\mathrm{R}(\omega)$ on the real-frequency axis to the propagator $G(\tau)$ on the imaginary-time axis:
\begin{equation}
    G(\tau) \equiv - \langle T_\tau A(\tau) B(0) \rangle  = -\int \dd\omega\; K(\tau,\omega)\,\rho(\omega),
    \label{eq:anacont}
\end{equation}
where $K$ is an integral kernel mediating the transition (cf.~Sec.~\ref{sec:arch}):
\begin{equation}
    K(\tau, \omega)
= \frac{\exp(-\tau\omega)}{\exp(-\beta\omega) \pm 1} \Theta(\wmax - |\omega|),
    \label{eq:K}
\end{equation}
$T_\tau$ is the time-ordering operator, $\wmax$ is a UV cutoff (upper bound to the bandwidth), and
$0 < \tau < \beta$ is imaginary time.
This information loss is epitomized by the singular value expansion (SVE)~\cite{Hansen:SVE} of the kernel $K$~\cite{Bryan90,Otsuki17}:
\begin{equation}
    K(\tau, \omega) = \sum_{l=0}^\infty U_l(\tau)\,S_l\,V_l(\omega),
    \label{eq:SVE}
\end{equation}
where $\{U_l\}$ are the left-singular functions, an orthonormal system on the imaginary-time axis,
and $\{V_l\}$ are the right-singular functions, an orthonormal system on the real-frequency axis.  The
amount of information retained in the transformation from $V_l$ to $U_l$ 
is encoded in the associated (scaled) singular value $S_l/S_0$. Crucially, $S_l$ decays at least exponentially quickly, $\log(S_l) \sim -l/\log(\beta\wmax)$.
This information loss on the other hand allows the imaginary-time propagator to be compressed
by storing the expansion coefficients of the left-singular functions (``IR basis functions'') $U_l$~\cite{Shinaoka17:compressing}:%
\begin{subequations}%
\begin{equation}
    G(\tau) = \sum_{l=0}^{L-1} U_l(\tau)\,G_l + \epsilon_L(\tau),
    \label{eq:gtauir}
\end{equation}
where $G_l = -S_l \int\dd\omega\,V_l(\omega) \rho(\omega)$ are the expansion coefficients and $\epsilon_L$ is an error term which vanishes exponentially quickly, $\epsilon_L \sim S_L/S_0$.  Its Fourier
transform is given by
\begin{equation}
    \hat G(\iw) = \int_0^\beta \dd\tau\,\ee^{\iw\tau} G(\tau) = \sum_{l=0}^{L-1} \hat U_l(\iw)\,G_l + \hat\epsilon_L(\iw),
    \label{eq:gwir}
\end{equation}%
\label{eqs:ireval}%
\end{subequations}%
where $\iw=\frac{\ii\pi}\beta(2n+\zeta)$ is a Matsubara frequency, $\zeta=0$/1 for bosons/fermions, and $\hat \cdot$ denotes the Fourier transform.
The singular value construction means that the IR basis is
(a) {\em optimal} in terms of compactness:\footnote{The truncated IR expansion minimizes $||\epsilon_L||$ in the $L_2$-norm sense if no additional information, i.e., a flat prior, for $\rho(\omega)$ is used.} for, e.g., $\beta\wmax < 10^8$, no more than 200 coefficients must be stored to obtain full double precision accuracy; 
(b) orthonormal; and
(c) unique, thereby providing a robust and compact storage format. 

To achieve (ii) reconstruction, we note that the ``po\-ly\-nomial-like'' properties of $U_l$~\cite{Karlin68,Wallerberger21:BSE} guarantee
that there exists~\cite{Rokhlin96,Shinaoka21:note} a sparse set of $\bigO(L)$ times $\TT=\{\tau_i\}$ and frequencies 
$\WW=\{\iw_n\}$ from which we can robustly infer the coefficients~\cite{Li20}.  \sparseir{} solves the
following ordinary least-squares problems:%
\begin{subequations}%
\begin{align}
    G_l &= \underset{\{G_l\}}{\operatorname{arg\,min}} \sum_{\tau\in\TT} \Big| G(\tau) - \sum_{l=0}^{L-1} U_l(\tau)\,G_l \Big|^2,
    \label{eq:gtaufit}\\
    G_l &= \underset{\{G_l\}}{\operatorname{arg\,min}} \sum_{\iw\in\WW} \Big| \hat G(\iw) - \sum_{l=0}^{L-1} \hat U_l(\iw)\,G_l \Big|^2.
    \label{eq:giwfit}
\end{align}%
\label{eqs:irfit}%
\end{subequations}
Given a sensible choice for the sampling points, Eqs.~(\ref{eqs:ireval}) and (\ref{eqs:irfit}) now allow us
to move between sparse imaginary-time and frequency grids and compressed representations without any significant
loss of precision~\cite{Li20}.

% ======================
\section{Example usage}
\label{sec:examples}
% ======================
As a simple example, let us perform self-consistent second-order perturbation theory
for the single impurity Anderson model at finite temperature.
Its Hamiltonian is given by
\begin{equation}
\begin{split}
H &= -\mu(\cdag_\uparrow\cee_\uparrow + \cdag_\downarrow\cee_\downarrow) + U \cdag_\uparrow \cdag_\downarrow \cee_\downarrow \cee_\uparrow \\
&\hphantom=
+ \sum_{p\sigma} \big(V_{p\sigma} \fdag_{p\sigma} \cee_\sigma + V^*_{p\sigma} \cdag_\sigma\eff_{p\sigma}\big)
+ \sum_{p\sigma} \epsilon_p \fdag_{p\sigma} \eff_{p\sigma},
\end{split}
    \label{eq:AIM}
\end{equation}
where $U$ is the electron interaction strength, $\mu$ is the chemical potential, $\cee_\sigma$ annihilates an electron on the impurity,
$\eff_{p\sigma}$ annihilates an electron in the bath, $\dagger$ denotes the Hermitian conjugate, $p\in\mathbb R$ is bath momentum, and $\sigma\in\{\uparrow, \downarrow\}$ is spin. The hybridization strength $V_{p\sigma}$ and
bath energies $\epsilon_p$ are chosen such that the non-interacting density of states is semi-elliptic
with a half-bandwidth of one, $\rho_0(\omega) = \frac2\pi\sqrt{1-\omega^2}$, $U=1.2$, $\beta=10$, and the system is half-filled, $\mu = U/2$.

\begin{figure}
%...,....1....,....2....,....3....,....4....,.
\begin{lstlisting}[language=Python]
import sparse_ir as ir, numpy as np
basis = ir.FiniteTempBasis('F', 10, 8, 1e-6)
U = 1.2
def rho0w(w):
  return np.sqrt(1-w.clip(-1,1)**2) * 2/np.pi
rho0l = basis.v.overlap(rho0w)
G0l = -basis.s * rho0l
Gl_prev = 0
Gl = G0l
stau = ir.TauSampling(basis)
siw = ir.MatsubaraSampling(basis)
while np.linalg.norm(Gl - Gl_prev) > 1e-6:
  Gl_prev = Gl
  Gtau = stau.evaluate(Gl)
  Sigmatau = U**2 * Gtau**3
  Sigmal = stau.fit(Sigmatau)
  Sigmaiw = siw.evaluate(Sigmal)
  G0iw = siw.evaluate(G0l)
  Giw = 1/(1/G0iw - Sigmaiw)
  Gl = siw.fit(Giw)
\end{lstlisting}
%...,....1....,....2....,....3....,....4....,.
\caption{Self-consistent second-order perturbation theory for a 
single-impurity Anderson model (\ref{eq:AIM}) with a semi-elliptic
density of states and $U=1/2$ at half filling and $\beta=10$
using \sparseir.}
\label{lst:ex}
\end{figure}

We present the associated algorithm in Fig.~\ref{lst:ex}.
First, we construct the IR basis for fermions and $\beta=10$, intuit that $\wmax=8$ is larger than
the interacting bandwidth and content ourselves with an accuracy of $\epsilon=10^{-6}$ (line 2).
We then compute the basis coefficients as $\rho_{0,l} = \int\dd\omega\,V_l(\omega) \rho_0(\omega)$
(line 6). The non-interacting propagator $G_{0,l} = -S_l \rho_{0,l}$ (line 7)
serves as initial guess for $G_l$ (line 9).
We then construct the grids and matrices for sparse sampling (lines 10, 11), after which
we enter the self-consistency loop (line 12):
At half filling, the second-order self-energy is simply
\begin{equation}
    \Sigma(\tau) = U^2 G^3(\tau)
    \label{eq:gf2}
\end{equation} (line 15).
We construct this object at the sampling points $\{\tau_i\}$ by first expanding $G_l$
(line 14). The dynamical part of the self-energy is propagator-like, so it
can be modeled by the IR basis (the Hartree and Fock term, if present, needs to be
handled separately). The Dyson equation
\begin{equation}
    \hat G^{-1}(\iw) = \hat G_0^{-1}(\iw) - \hat\Sigma(\iw)
    \label{eq:dyson}
\end{equation} (line 19)
is then solved by expanding both $G_0$ and $\Sigma$ on the sparse set of frequencies (lines 17, 18).
To complete the loop, the IR coefficients for $G$ are then updated (line 20).  We converge if the deviation between subsequent iterations (line 13) is consistent with the basis accuracy (line 12).

\begin{figure}[t]
\centering
\includegraphics[width=\columnwidth]{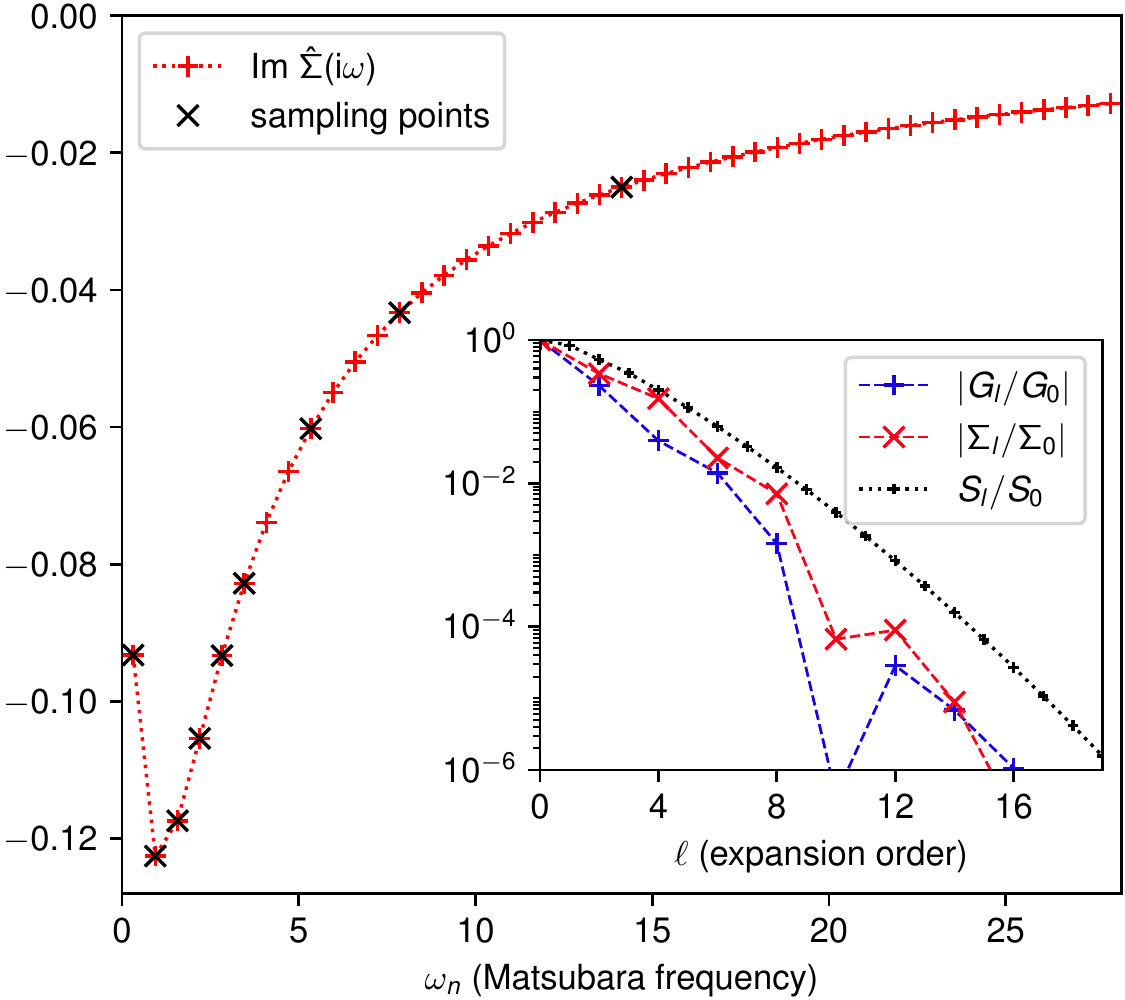}
\caption{Imaginary part of the Matsubara self-energy $\hat\Sigma(\iw)$ for
the GF(2) calculation in Fig.~\ref{lst:ex}.  Black crosses mark the location
of sampling points $\WW$ on which the Dyson equation (\ref{eq:dyson}) is solved
and from which the full signal is reconstructed.
Inset: normalized IR expansion coefficients of $G$ (plusses) and $\Sigma - \Sigma_\mathrm{HF}$
(crosses) and singular values (dots) for comparison. Lines are guides for the eye.
}
\label{fig:gf2}
\end{figure}

The resulting self-energy $\hat\Sigma(\iw)$ on the Matsubara axis is presented in Fig.~\ref{fig:gf2} (only the imaginary part is plotted, since the real part is merely a constant $U/2$ at half filling).  Instead of a dense mesh (plusses), the Dyson equation has to be solved only on the sampling points (crosses).  Since the IR coefficients for both the Green's function and the self-energy are guaranteed to decay quickly (see inset), this is enough to reconstruct the functions everywhere with the given accuracy bound of $\epsilon=10^{-6}$.  We note that this bound and the UV cutoff $\wmax$ are the only discretization parameters we need to supply.

The code in Fig.~\ref{lst:ex} is short, simple---no explicit Fourier transforms or models
are required---yet guarantees the given accuracy goal.  Extending the approximation to $\Sigma=GW$
would require only the addition of a bosonic basis, the construction of the RPA diagram,
$\Pi(\tau) = G^2(\tau)$, and solving the Bethe--Sal\-pe\-ter equation, $\hat W(\iw) = U + U \hat\Pi(\iw) \hat W(\iw)$,
where again sparse grids and transformations can be used.
%\HS{[We have not mentioned the bosonic basis yet]}

In addition to this example, \sparseir{} ships a set of tutorials~\cite{sparse-ir-tutorial}, demonstrating the use of the Python, Julia, and Fortran libraries in typical many-body calculations.
Each tutorial contains a short description of the underlying many-body theory as well as sample code utilizing \sparseir{} and its expected output.  Currently, we include tutorials on:
(a) the GF(2) and \textit{GW} approximation~\cite{PhysRev.139.A796,Aryasetiawan_1998},
(b) fluctuation exchange (FLEX) \cite{Bickers89a,Niklas2020,witt2022doping},
(c) the two-particle self-consistent (TPSC) approximation \cite{Vilk1997,Tremblay2012},
(d) Eliashberg theory for the Holstein--Hubbard model 
\cite{Eliashberg60,Scalapino_book,PhysRevLett.69.1600,kaga2022eliashberg},
(e) the Lichtenstein formula~\cite{Liechtenstein_1984},
(f) calculation of the orbital magnetic susceptibility \cite{Fukuyama, gomez-santos,raoux,piechon,ogata-fukuyama,matsuura-ogata,ogata2,ogata3}, and
(g) numerical analytic continuation based on the SpM method~\cite{Otsuki17}.

% ==================================
\section{Architecture and features}
\label{sec:arch}
% ==================================
\begin{figure*}
    \centering
    \includegraphics[scale=.7]{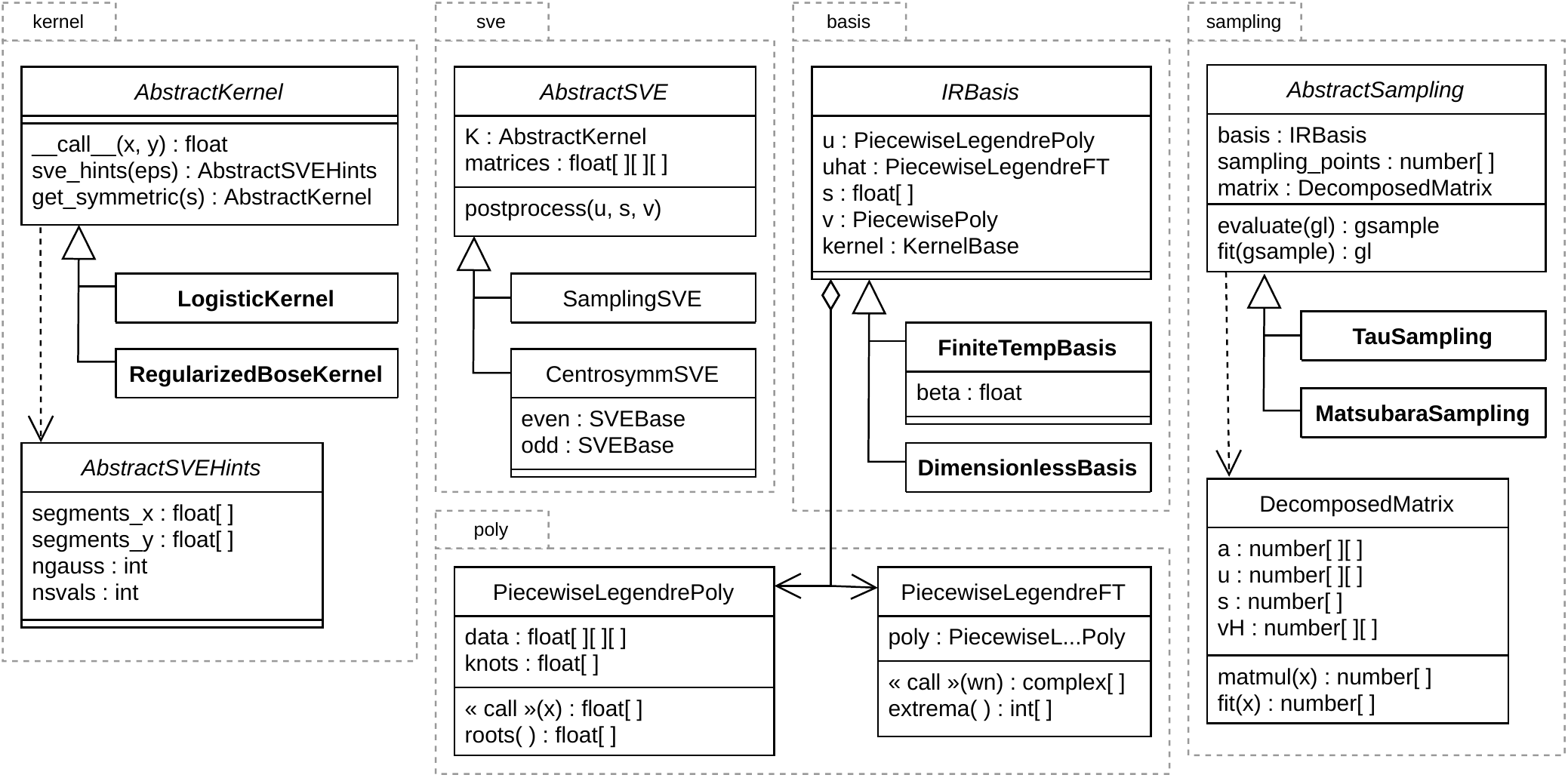}
    \caption{Simplified UML class diagram of the core of \sparseir{}. Classes with names in
             boldface are available from the top namespace.}
    \label{fig:uml}
\end{figure*}
The main functions of the library are (a) construction and handling of the kernel (\ref{eq:K}), (b) performing the singular value expansion (\ref{eq:SVE}), (c) storage and evaluation of the IR basis functions (\ref{eqs:ireval}), and (d) construction of the sampling points and
solution of the fitting problem (\ref{eqs:irfit}).  The \sparseir{} package was split along these lines into
modules, see Fig.~\ref{fig:uml}, which we will briefly describe in the following.

\paragraph{Kernel}
Two kernels are packaged with \sparseir:%
\begin{subequations}%
\begin{align}
    K^\mathrm{log}_\Lambda(x, y) &= \frac{\exp(-\tfrac\Lambda2 (x + 1)y)}{1 + \exp(-\Lambda y)} \Theta(1-|y|),\label{eq:Klog} \\
    K^\mathrm{RB}_\Lambda(x, y) &= \frac{y \exp(-\tfrac\Lambda2 (x + 1)y)}{1 - \exp(-\Lambda y)}
    \Theta(1-|y|)\label{eq:KRB}.
\end{align}%
\label{eqs:Kpredef}%
\end{subequations}%
Kernels are
expressed in terms of dimensionless variables $x$ and $y$ in the interval $[-1, 1]$, where $\tau = \frac\beta2(x+1)$ and $\omega = \wmax y$.  Instead of parametrization by both inverse temperature $\beta$ and UV cutoff frequency $\wmax$,
this allows one to consider only a single scale parameter $\Lambda = \beta\wmax$.

The logistic kernel (\ref{eq:Klog}) is the default kernel used for both fermionic and bosonic propagators
for simplicity: even though it is the analytic continuation kernel for fermions, it can also be used to compactly model bosonic propagators~\cite{Meyer07,Itou21,DLR21}.  The regularized bosonic kernel (\ref{eq:KRB}) is common in numerical analytic continuation
of bosonic functions~\cite{Jarrell96:AnaCont,Motoyama22} and is used by the {\em irbasis} library for bosonic propagators.  Thermal contributions to the susceptibility, $\chi(\omega=0)$, can be modelled by augmenting the basis~\cite{Wallerberger21:BSE}. User-defined kernels may be added.

Each kernel $K$ can be evaluated by supplying $x,y$, however care must be taken not to lose precision around $x=\pm1$: in addition to $x$ we use $x_\pm \coloneqq 1 \pm x$ to full precision to avoid cancellation in the enumerators of Eqs.~(\ref{eqs:Kpredef}).

% ---------------------------------
\paragraph{Piecewise polynomials}
% ---------------------------------
To represent the IR basis functions (\ref{eq:SVE}), we employ piecewise Legendre polynomials:
\begin{equation}
P_{nq}(x; \{x_m\}) \coloneqq \sqrt{\frac 1{\Delta x_n}} P_q\!\left( \frac{x - \bar x_n}{\Delta x_n} \right) 
\Theta(\Delta x_n - |x - \bar x_n|),
\end{equation}
where $x_0<x_1<\cdots<x_N$ are the segment edges, $\Delta x_n \coloneqq \frac12(x_n - x_{n-1})$, $\bar x_n \coloneqq \frac12(x_n + x_{n-1})$, and $P_q$ denotes the $q$-th Legendre
polynomial.

Given suitable discretizations of the axes, $\{x_n\}$ and $\{y_n\}$, as well as a Legendre order $Q$,
the left and right IR basis functions can then be expanded as follows:%
\begin{subequations}%
\begin{align}%
    u_l(x) &\approx \sum_{n=1}^N \sum_{q=0}^Q u_{lnq} P_{nq}(x; \{x_0, \ldots, x_N\}), \\
    v_l(y) &\approx \sum_{n'=1}^{N'} \sum_{q'=0}^Q v_{ln'q'} P_{n'q'}(y; \{y_0, \ldots, y_{N'}\}),
\end{align}%
\label{eqs:uvexpand}%
\end{subequations}
where $u_{lnq}$ and $v_{ln'q'}$ are expansion coefficients. 

Legendre polynomials have the advantage that their Fourier transform is given analytically~\cite{Boehnke11}:
\begin{equation}
\begin{split}
\hat P_{nq}(\pm\omega; \{x_m\}) &\coloneqq \int_{x_0}^{x_N} \dd{x} \ee^{\pm\iw x} P_{nq}(x; \{x_m\}) \\
&= 2 \sqrt{\Delta x_n} \ee^{\pm\iw\bar x_n} (\pm\ii)^q j_q(\omega\Delta x_n),
\end{split}
\label{eq:Phat}
\end{equation}
where $\omega\ge 0$ is a frequency and $j_q(x)$ is the $q$-th spherical Bessel function.  Thus, no
numerical integration is necessary, though $\exp(\iw x)$ must be analytically mapped back to small $\iw x$
to avoid cancellation.

% ---------------------------------
\paragraph{Singular value expansion (SVE)}
% ---------------------------------
Given the discretization outlined above, we can relate the SVE (\ref{eq:SVE}) needed
for constructing the IR basis to the singular value decomposition (SVD) of the following $(NQ)\times(N'Q)$ matrix~\cite{Hansen:SVE,Shinaoka17:compressing}:
\begin{equation}
\begin{split}
  &A_{nq,n'q'} = \sqrt{(q + \tfrac12)(q' + \tfrac12)}\\
  &\quad\times \iint \dd{x} \dd{y} P_{nq}(x, \{x_m\}) P_{n'q'}(y, \{y_m\}) K(x, y),
\end{split}
\label{eq:svd}
\end{equation}
where the singular values of $A$ are equal to the singular values $s_l$ of 
the kernel, and the left and right singular vectors are the
(scaled) expansion coefficients of $u_l(x)$ and $v_l(y)$, respectively (\ref{eqs:uvexpand}). $L$ is chosen such that
$s_L<\epsilon s_0$, where $\epsilon$ is the desired accuracy of the basis.
In practice, we approximate
the integral (\ref{eq:svd}) by the associated Gauss--Legendre rule and rewrite the problem
as equation for the Gauss nodes~\cite{Rokhlin96,Hansen:SVE,DLR21}.
As the kernels (\ref{eqs:Kpredef}) are all centrosymmetric, $K(x,y)=K(-x,-y)$, the 
SVE problem is block-diagonalized for a four-fold speedup~\cite{Chikano18}.

We empirically find that choosing $\{x_m\}$ and $\{y_m\}$ close to the extrema of the 
highest-order basis functions, $u_{L-1}(x)$ and $v_{L-1}(y)$, respectively, to provide
an excellent discretization, only necessitating $Q=16$ for $\epsilon = 10^{-16}$.
Since computing the basis functions requires solving the SVE, each kernel maintains
approximations to $\{x_m\}$ and $\{y_m\}$ as hints.
As only a fraction $1/Q$ of the singular values of Eq.~(\ref{eq:svd}) are needed,
we use a truncated SVD algorithm (rank-revealing $QR$ decomposition followed by 
two-sided Jacobi rotations~\cite{GolubVanLoan}) at the cost of $\bigO(N'^2 N Q^3)$.

In order to guarantee an accuracy of $\epsilon$ for both 
singular values and basis functions, one has to compute the SVE with a machine precision
of $\epsilon^2$~\cite{GolubVanLoan}.  Thus we compute the SVD in standard double precision for $\epsilon \ge 10^{-8}$
and quadruple precision otherwise.  For the latter we have developed the
{\em xprec} extension to numpy. Note that quadruple precision is only needed in the
SVE -- the basis functions are stored and evaluated in double precision.

%With these algorithmic improvements, \sparseir{} allows to compute the IR basis
%functions on the fly rather than

% ---------------------------------
\paragraph{Sampling}
% ---------------------------------
The sampling times $\{\tau_i\}$ and frequencies $\{\iw_n\}$ are chosen such that
the highest-order basis functions, $U_{L-1}(\tau_i)$ and $\hat U_{L-1}(\iw_n)$,
respectively, are locally extremal.  To optimize conditioning,  $\tau_1$ and $\tau_L$ 
are moved from $\pm1$ to the midpoint of between the $\pm1$ and the closest root
of $U_{L-1}$.  Sampling in frequency is conditioned somewhat worse due to the discrete
nature of the frequency axis, which is why $\{\iw_n\}$ are augmented 
by four additional frequencies.

With the sampling points chosen, sparse sampling now involves transitioning between
IR basis coefficients and the value at the sampling points.  For evaluation (\ref{eqs:ireval})
at the sampling points we multiply with precomputed matrices, $F_{il} \coloneqq U_l(\tau_i)$ and
$\hat F_{nl} \coloneqq \hat U_l(\iw_n)$, respectively, at a cost of $\bigO(L^2)$.  
For fitting the IR coefficients, we need to solve the least-squares problems (\ref{eqs:irfit}).
However, multiplying with a precomputed pseudoinverse can lead to loss of backward stability~\cite{Wilkinson63}, and we observe this in the case of basis augmentation.  
Instead, we precompute and store the SVD of $F$ and $\hat F$ and
construct the pseudoinverse on the fly, again at a cost of $\bigO(L^2)$.  

% -------------------------
\paragraph{Julia and Fortran libraries}
% -------------------------
This software package includes Julia~\cite{sparse-ir-jl} and Fortran~\cite{sparse-ir-fortran} libraries.
The Julia library implements the full set of functionalities of the Python library with a similar interface.
The Fortran library implements only their subset required for its use in \textit{ab initio} programs: The Fortran library uses the tabulated values of the IR basis functions computed by the Python library.
The Fortran interface is fully compatible with the Fortran95 standard and has no additional external dependencies.
More detailed descriptions can be found in readme files of the repositories and the tutorials described below.

% ======================================
\section{Impact and outlook}
\label{sec:impact}
% ======================================
We expect that the library will be widely used in many-body and \textit{ab initio} calculations based on diagrammatic theories such as \textit{GW} and quantum embedding theories such as the dynamical mean-field theory and its extensions.
The computational complexity of diagrammatic calculations based on these technologies grows slower than any power law with respect to the inverse temperature.
This makes these technologies particularly efficient and useful in studying systems with a large bandwidth at low temperatures. 
The library will make new studies for understanding the low-temperature properties of solids and molecules
feasible.

To facilitate its application to various fields, the library supports languages popular in many different areas (Python and Julia for prototyping, Fortran, C, and C++ for existing {\em ab-initio} codes.)
The library is shipped with many self-contained tutorials on specific topics in different fields of physics.
%Python is one of the most popular languages in scientific computing, while Julia has been attracting growing interest in the community.
%Fortran is widely used, especially in many \textit{ab initio} programs.
%The Fortran interface can also be called from C and C++.

% ======================================
\section{Conclusions}
\label{sec:conclusions}
% ======================================
We present intermediate representation (IR) and sparse sampling for efficient many-body and \textit{ab initio} calculations based on imaginary-time propagators.
These methods are implemented in Python/Julia/Fortran libraries to allow researchers in a large community of many-body physics and \textit{ab initio} calculations to use them.

\section*{Conflict of interest}
We wish to confirm that there are no known conflicts of interest associated with this publication and there has been no significant financial support for this work that could have influenced its outcome.

\section*{Acknowledgements}
\label{sec:acknow}

MW was supported by the FWF through project P30997.
NW acknowledges funding by the Cluster of Excellence `CUI: Advanced Imaging of Matter' of the DFG (EXC 2056 - project ID 390715994) and support by the DFG research unit QUAST FOR5249 (project DFG WE 5342/8-1).
RS, FK and HS were supported by JST, PRESTO Grant No. JPMJPR2012.
HS was supported by JSPS KAKENHI Grants No. 21H01041 and No. 21H01003.
SH was supported by No. JP21K03459.
TK was supported by JSPS KAKENHI Grants No. 21H01003, 21H04437, and 22K03447.
SO was supported by JSPS KAKENHI Grants No. 18H01162 and JSPS through the
Program for Leading Graduate Schools (MERIT). 
KN was supported by JSPS KAKENHI Grants No. JP21J23007.

\bibliographystyle{elsarticle-num}
\bibliography{main}
% Please add the reference to the software repository if DOI for software is available.

\end{document}